\documentclass[wssquare]{ws-procs975x65}            
\begin{document}
\title{Numerically reconstructing the geometry of the Universe from data}

\author{Hertzog L. Bester$^*$, Julien Larena and Nigel T. Bishop}

\address{Department of Mathematics, Rhodes University,\\
Grahamstown, 6139, South Africa\\
$^*$E-mail: g07b1135@campus.ru.ac.za\\
www.ru.ac.za}

\begin{abstract}
We give an outline of an algorithm designed to reconstruct the background cosmological metric within the class of spherically symmetric dust universes that may include a cosmological constant. Luminosity and age data are used to derive constraints on the geometry of the universe up to a redshift of $z = 1.75$. It is shown that simple radially inhomogeneous void models that are sometimes used as alternative explanations for the apparent acceleration of the late time Universe cannot be ruled out by these data alone. 
\end{abstract}


\bodymatter

\section{Overview}\label{sec:OV}
The \emph{observational cosmology programme} \cite{Ellis1985} aims to reconstruct the background geometry (equivalently the cosmological metric) directly from observations. Here we give a brief overview of a newly proposed algorithm \cite{bester2} that solves this problem for the class of spherically symmetric dust universes that may include a cosmological constant (henceforth $\Lambda$LTB models, see \cite{Valkenburg:2011tm} for example). The algorithm employs a non-parametric approach in which Gaussian process priors are used to fix the free functions of the model. These priors are, as far as possible, informed by observational data. The two free functions of the model are chosen to be the longitudinal expansion rate $H_\|(z)$ and the energy density of cold dark matter $\rho(z)$. Together with the value of the cosmological constant $\Lambda$, samples of these two functions completely specify a $\Lambda$LTB model. The geometry of the Universe can then be solved for numerically by using observational coordinates \cite{Ellis1985} to pose the Einstein field equations (EFE) as a characteristic initial value problem (CIVP) (see \cite{prd2010} and \cite{prd2012} for details). Once the form of the metric is known it can be used, in conjunction with the forms of the fluid variables $\rho$ and $u^a$, to compute any observable as a function of redshift $z$. This allows the current realisations of $H_\|(z)$, $\rho(z)$ and $\Lambda$ to be confronted with data. Inference is achieved by formulating a Bayesian model for the problem and implementing a modified random walk algorithm \cite{cotter2013} over the function space of $H_\|(z)$ and $\rho(z)$\footnote{Since the data considered do not constrain the value of the cosmological constant it is treated as a nuisance parameter.}. For convenience we refer to this random walk algorithm as an MCMC even though it is not a true Markov-Chain-Monte-Carlo sampler. This work is an extension of work initiated in \cite{bester1} and \cite{bester2} which could be consulted for further details. 

\section{Framework}
The metric in observational coordinates $x^a = [w,v,\theta,\phi]$ can be written as (see \cite{bester1,bester2} for further clarification of our notation)
\begin{equation}
ds^2 = -A(w,v)dw^2 + 2dw dv + D(w,v)^2 d\Omega^2.
\label{ObsMet}
\end{equation} 
Substituting \eqref{ObsMet} into the EFE in the form $R_{ab} + \Lambda g_{ab} = \kappa(T_{ab} - \frac{1}{2}Tg_{ab})$ results in
\begin{eqnarray}
D'' &=& -\frac{1}{2}\kappa D \rho (u_1)^2, \label{RNU} \\
\dot{D}' &=& \frac{1}{2D} \left[1- DD'A' - 2\dot{D}D' - A(D')^2 - ADD'' - \frac{1}{2}\kappa \rho D^2 - \Lambda D^2\right], \label{RW} \\
A'' &=& \kappa A(u_1)^2 \rho - 4\frac{\dot{D}'}{D} -2\frac{A'D'}{D} - 2\Lambda,  \label{WNU}\\
D(0) &=&  A'(0) = \dot{D}(0) = 0, ~~ A(0) = D'(0)=1, \label{ICs}
\end{eqnarray}
where a dot denotes the partial derivative w.r.t. $w$ and a prime the partial derivative w.r.t. $v$. The inputs required to solve this system are the functional forms of $\rho(v)$ and $u_1(v) = 1 + z(v)$, as well as the value of $\Lambda$. Clearly the $z(v)$ relation is required to convert between functions of the redshift $z$ and the null affine parameter $v$. This relation follows from projecting the null geodesic equation along the direction of propagation of the ray and is given by\footnote{Note that for positive $H_\|$ this relation is one to one and therefore uniquely invertible.}
\begin{equation}
v(z) = \int_0^{z_{\mbox{\tiny{max}}}} \frac{dz}{(1+z)^2 H_\|(z)}.
\label{nuz}
\end{equation}
Thus, given $H_\|(z), ~\rho(z)$ and $\Lambda$, we can solve for the metric components $D$ and $A$ on the current past lightcone (henceforth PLC0). Once the solution is known it can be used to compute any observable in terms of $D,~A,~\rho,~u$ and their derivatives. This is all that is required to perform inference on the input set $[H_\|,\rho,\Lambda]$. However, it is possible to obtain information about the state of the Universe at earlier times by evolving the system into the past. This is achieved using the conservation equations $\nabla_aT^{ab} = 0$ which, after some simplification, yield 
\begin{eqnarray}
\dot{u}_{1} &=& \frac{1}{2}\left[(\frac{1}{(u_1)^2} - A)(u_{1})' -A'u_1\right], \label{U1W}\\
\dot{\rho} &=& \rho\left(-\frac{(u_1)'}{u_1^{\ 3}} - 2\frac{\dot{D}}{D} + \frac{D'}{D}\left(\frac{1}{u_1^{\ 2}} - A\right)\right) + \frac{1}{2}\rho'\left(\frac{1}{u_1^{\ 2}} - A\right) \label{RHOW}
\end{eqnarray}
The above equations describe the evolution of the inputs from one past lightcone to the next and therefore allow us to find the solution in the interior of our PLC. Since the evolution of the Universe is significantly different for different values of $\Lambda$, such a solution provides a probe of the value of the cosmological constant. In fact the value of $\Lambda$ can be inferred directly from redshift drift data (see the discussion in \cite{bester2}). However, since such data are not currently available, we are only able to constrain the joint distribution of $H_\|$ and $\rho$. The value of $\Lambda$ is marginalised over by sampling it from a fairly generous flat prior distribution (see \cite{bester2} for details).\\
The most efficient and robust priors over $H_\|(z)$ and $\rho(z)$ result from performing Gaussian process (GP) regression (see \cite{rasmussen2006gaussian} for example) on the available\footnote{Since there are no model independent $\rho(z)$ data available the prior over $\rho(z)$ is actually set using mock data as explained in \cite{bester2}. Importantly these data are not used for inference.} data. Denoting the target vector by $x$, we construct the prior $\mu_0(x)$ assuming that these two functions are independent i.e.
\begin{equation}
x = \left(\begin{array}{c}
H_\| - \bar{H}_\| \\ 
\rho - \bar{\rho}
\end{array}\right), \quad \mbox{with} \quad \mu_0(x) \sim \mathcal{N} \left( 0, \left(\begin{array}{cc}
\Sigma_H & 0 \\ 
0 &  \Sigma_{\rho}
\end{array}\right) \right).
\end{equation}
The notation $\mathcal{N}(a,A)$ is used to denote a multivariate Gaussian distribution with mean $a$ and covariance matrix $A$. Thus $\Sigma_H$ and $\Sigma_\rho$ are the posterior covariance matrices for the GP's over $H_\|$ and $\rho$, respectively, and we have centred the target vector to zero by subtracting out the posterior GP mean functions $\bar{H}_\|$ and $\bar{\rho}$. Denoting the observables jointly by $y$, the Bayesian model for the problem takes the form
\begin{equation}
y = \mathcal{H}(x) + \epsilon, \quad \mbox{with} \quad \epsilon \sim \mathcal{N} \left( 0, \Sigma_y \right),
\end{equation}
where $\Sigma_y$ is the covariance matrix of the data, $\mathcal{H}$ is the hypothesis which can be identified with the system \eqref{RNU}-\eqref{RHOW} and $\epsilon$ is noise. The assumption that $\epsilon$ is jointly normally distributed determines the form of the likelihood function as a Chi-square distribution. Inference (see \cite{cotter2013} for a detailed discussion of the inference framework employed in this work) can be performed on $x$ by using the Radon-Nikodym derivative to express Bayes' law as
\begin{equation}
\frac{d\mu}{d\mu_0}(x) \propto L(x), \quad \mbox{with} \quad L(x) = \exp(-\chi^2(x)),
\label{RNBayThe}
\end{equation} 
where $L$ is the likelihood. The rate of convergence of the MCMC can be accelerated by using carefully constructed proposal distributions which exactly preserve the form of $\mu_0(x)$ as $L(x) \rightarrow 0$. We have used the preconditioned Crank-Nicolson proposal
\begin{equation}
\tilde{x}_{(k)} = \sqrt{(1-\beta^2)}x_{(k)} + \beta \delta, \quad \mbox{with} \quad \delta \sim \mu_0(x),
\label{pCNPro}
\end{equation}
in which $\beta \in [0,1]$ is a constant used to control the acceptance rate of the MCMC and the step in the chain is indicated by a subscript in round braces. Defining the acceptance probability as
\begin{equation}
a(x,\tilde{x}) = \min\left(1,\exp\left(\chi^2(x) - \chi^2(\tilde{x})\right)\right),
\label{AccPro}
\end{equation}
ensures that the method does not reject in the case when $\Phi = 0$ but rather accepts with probability one. 

\section{Results}
\begin{figure}[h]%
\hspace*{-1.1in}
\includegraphics[width=1.4\textwidth]{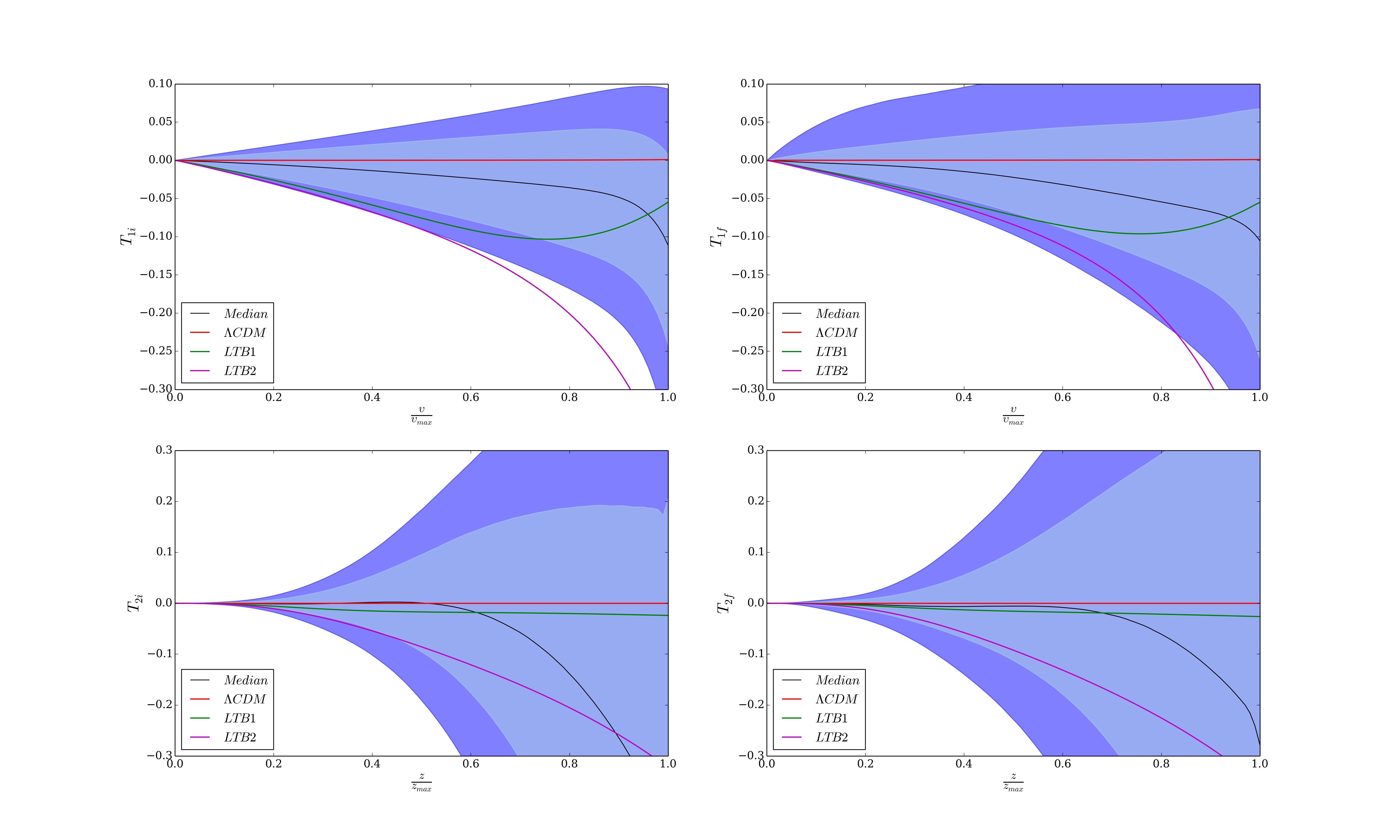}
\caption{Tests of the Copernican principle. Top) The quantity $T_1$ on the initial (left) and final (right) PLC. Bottom) The quantity $T_2$ on the initial (left) and final (right) PLC}
\label{fig:CP}
\end{figure}

For the current application we have used luminosity and age data to perform inference. In particular we convert the Union 2.1 compilation \cite{Suzuki:2011hu} to angular diameter distance data assuming distance duality (note we have not marginalised over $H_0$ for this data set). The cosmic chronometer approximation \cite{Moresco:2012by} is used to get longitudinal expansion rate data. Having smoothed these data using Gaussian process regression, we can draw function realisations of $H_\|(z)$ and $\rho(z)$ and therefore perform the MCMC described above. 
\def\figsubcap#1{\par\noindent\centering\footnotesize(#1)}
\begin{figure}[h]%
\begin{center}
 \parbox{2.41in}{\includegraphics[width=2.5in]{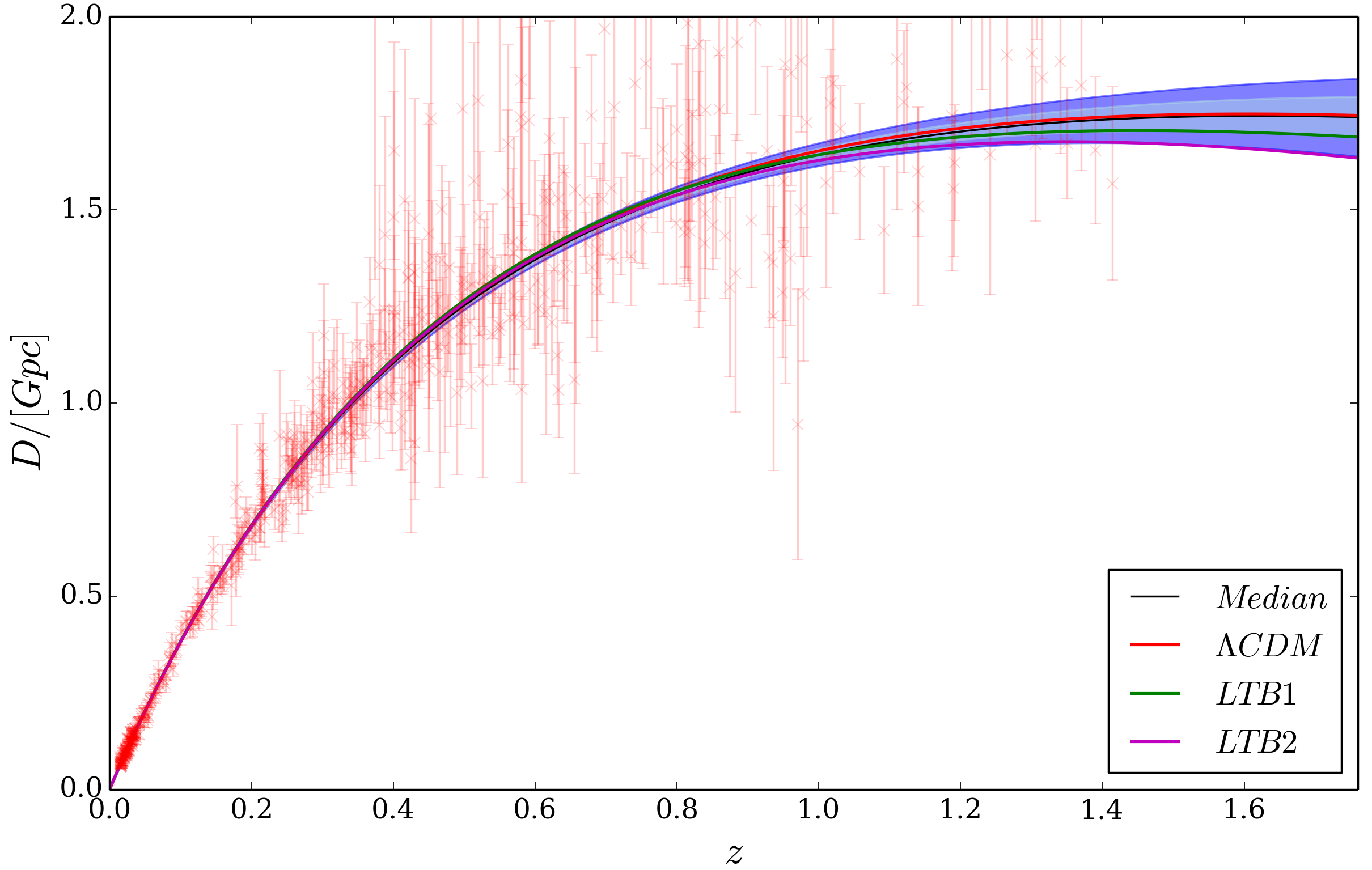}\figsubcap{a}}
 \hspace*{2pt}
 \parbox{2.41in}{\includegraphics[width=2.5in]{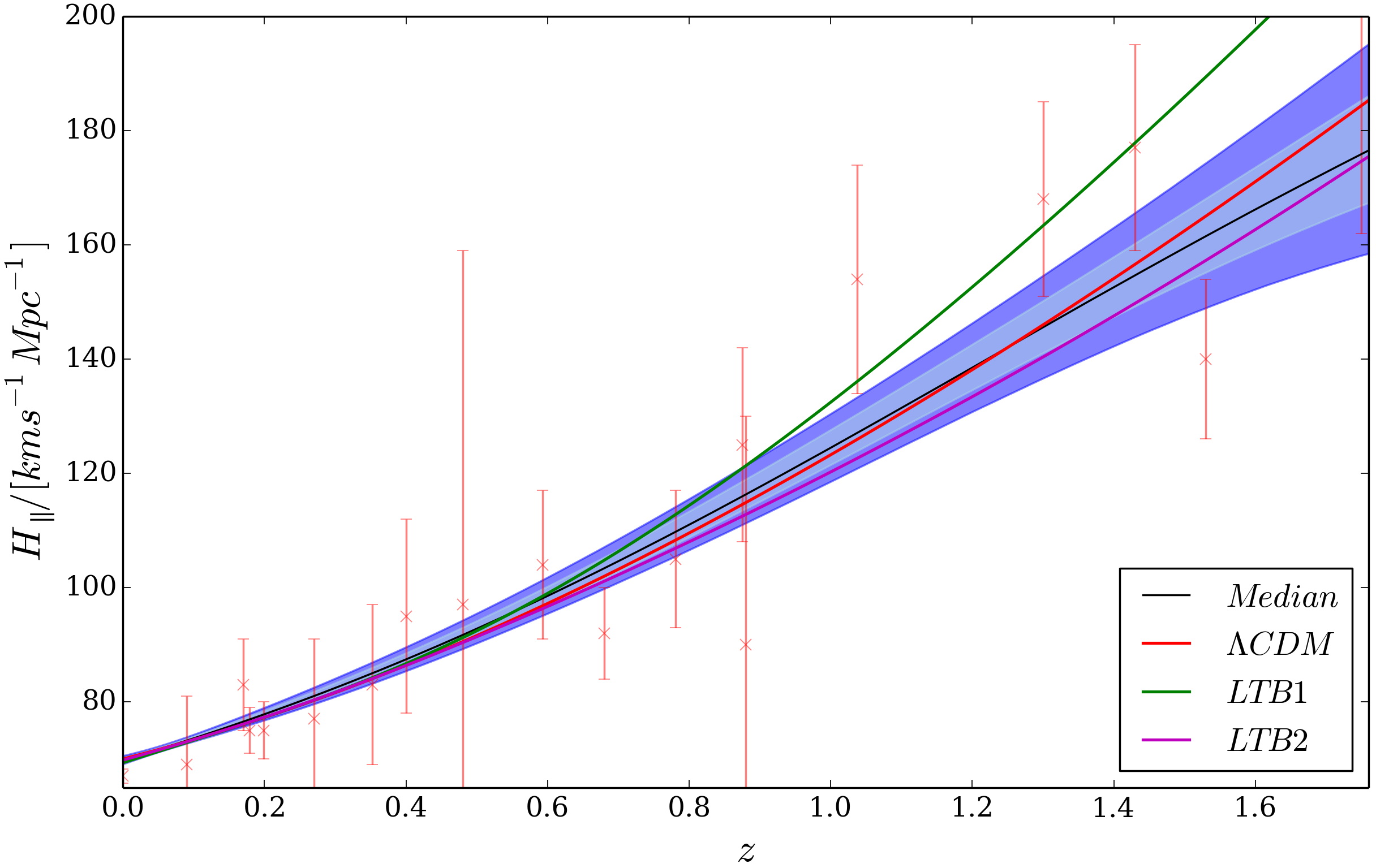}\figsubcap{b}}
 \caption{The data on the PLC0. (a) Angular diameter distance. (b) Longitudinal expansion rate.}
\label{fig:PLC0}
\end{center}
\end{figure}
The numerical error of the integration scheme is set to $10^{-5}$ and for each sample we compute two dimensionless consistency relations, both of which should evaluate to zero if the background universe is statistically homogeneous and isotropic. The first consistency relation is related to the matter shear and can be written as
\begin{equation}
T_1 = 1 - \frac{H_{\perp}}{H_\|},
\label{sheartest}
\end{equation}
where $H_\perp$ is the expansion in the direction transverse to the line of sight. The second consistency relation tests whether the dimensionless curvature parameter $\Omega_{K}$ is constant with redshift and can be written as (see \cite{Clarkson:2007pz})
\begin{equation}
T_2 = 1 + H_{\|}^2\left[u^2(DD_{,zz} - D_{,z}^2) - D^2\right] + uH_{\|}H_{\| ,z}D\left[uD_{,z} + D\right]. \label{curvetest}
\end{equation}
For comparison we also show these relations for both the constrained (i.e. $t_B(r) = 0$) and general best fit LTB models parametrised as in \cite{GarciaBellido2008nz}. Figure \ref{fig:PLC0} shows the data used for inference as well as the posterior distributions of $D(z)$ and $H_\|(z)$. Clearly the $\Lambda$CDM and unconstrained LTB (labelled LTB2 in the figure) models fall well within the reconstructed contours. The constrained LTB model (labelled LTB1 in the figure) has difficulty fitting both $D(z)$ and $H_\|(z)$ data simultaneously. In Figure \ref{fig:CP} we show the quantities $T_1$ and $T_2$ corresponding to the initial data shown in Figure \ref{fig:PLC0}. It is interesting to note that, while both the $\Lambda$CDM and constrained LTB models fall within the 2-$\sigma$ contours of the reconstructed distributions of $T_1$ and $T_2$, the unconstrained model seems to be disfavoured by the quantity $T_1$ at the 2-$\sigma$ confidence level. However, it has to be kept in mind that only the best fit models are shown in these figures. For the model to be excluded we would have to confirm that the confidence intervals of the model do not overlap with those reconstructed directly from the data. Doing so reveals that the current data cannot exclude either LTB model, even at the 1-$\sigma$ level. 
\vspace{-1cm}
\vspace{-1cm}
\section{Conclusion}
We have presented an algorithm capable of reconstructing the background geometry of the Universe directly from data. The algorithm does not presuppose a parametrisation for the input functions. Instead it uses Gaussian process regression to smooth $H_\|(z)$ and $\rho(z)$ data, both of which are in principle directly observable. This allows the priors to be informed by data data. It was shown that, with this direct non-parametric approach, current data are not sufficient to rule out simple radially inhomogeneous models that are sometimes used as alternative explanations of the apparent acceleration of the late-time Universe.  

\section*{Acknowledgements}
The financial assistence of the South African Square Kilometre project (SA SKA) towards HLB's research is hereby acknowledged. Opinions expressed and conclusions arrived at are those of the author and are not necessarily to be attributed to the SKA SA. NTB and JL are supported by the National Research Foundation (South Africa).

\bibliographystyle{ws-procs975x65}
\bibliography{HL_Bester.bib}

\end{document}